\documentclass[aps,twocolumn,pra,showpacs,superscriptaddress]{revtex4-1}

\usepackage{amsfonts,amsthm}
\usepackage{subfigure}
\usepackage{amssymb}
\usepackage{amscd}
\usepackage{amsmath}    
\usepackage{multirow}
\usepackage{graphicx}
\usepackage{dcolumn}
\usepackage{bm}

\newcommand{\bra}[1]{\mbox{$\left\langle #1 \right|$}}
\newcommand{\ket}[1]{\mbox{$\left| #1 \right\rangle$}}
\newcommand{\braket}[2]{\mbox{$\left\langle #1 | #2 \right\rangle$}}

\begin{document}
\title{Discrete-phase-randomized measurement-device-independent quantum key distribution}
\author{Zhu Cao}
\email{caozhu@ecust.edu.cn}
\address{Key Laboratory of Advanced Control and Optimization for Chemical Processes of Ministry of Education, East China University of Science and Technology, Shanghai 200237, China}
\address{Shanghai Institute of Intelligent Science and Technology, Tongji University, Shanghai 200092, China}

\begin{abstract}
Measurement-device-independent quantum key distribution removes all detector-side attacks in quantum cryptography, and in the meantime doubles the secure distance. The source side, however, is still vulnerable to various attacks. In particular, the continuous phase randomization assumption on the source side is normally not fulfilled in experimental implementation and may potentially open a loophole. In this work, we first show that indeed there are loopholes for imperfect phase randomization in measurement-device-independent quantum key distribution by providing a concrete attack. Then we propose a discrete-phase-randomized measurement-device-independent quantum key distribution protocol as a solution to close this source-side loophole.
\end{abstract}

\vspace{2pc}

\maketitle

\section{Introduction}
Quantum key distribution (QKD) provides an information-theoretically secure method to distribute identical keys between two parties, and is hence one of the most important ingredients in information-theoretically secure communication. The first QKD protocol was developed by Bennett and Brassard in 1984~\cite {Bennett:BB84:1984}, which consisted of two sides, a source side and a detector side. We refer to this protocol as BB84 hereafter. The security of BB84, however, relies on a few idealized assumptions. These assumptions are often violated in practice, which allows attacks mostly on the detector side. Measurement-device-independent (MDI) QKD is hence developed to close all loopholes on the detector side~\cite{Lo:MDIQKD:2012}. To achieve higher security, it is ideal to also close loopholes on the source side in MDI-QKD.

In an idealized MDI-QKD, each of the two parties, called Alice and Bob, provide single photons in the eigenstates of the rectilinear basis or the diagonal bases. The measurement device performs a Bell measurement on Alice's and Bob's signals. It can be shown that if both Alice and Bob choose the rectilinear basis, they can recover identical keys based on the Bell measurement outcomes. The events that Alice and Bob choose different bases are discarded. It was shown that the security of this protocol can be proved by treating the protocol as the time-reversed version of an entanglement-based QKD \cite{Ekert:QKD:1991}.

In a practical scenario, single photon sources are not available. Instead, a phase-randomized weak coherent laser is often utilized to approximate a single photon source. However, continuous phase randomization is impossible to be realized experimentally. Heuristically, the laser is turned off and then on again to approximate phase randomization, but there is no theoretical guarantee that this can provide perfect phase randomization. Indeed there is evidence that this method is far from perfect phase randomization~\cite {Xu:QRNG:2012}. 

Failure in phase randomization can yield the QKD system insecure with respect to the original security analysis. In a related work \cite{cao2015discrete}, it was shown that the phase randomization loophole in BB84  can be closed by using discrete phase randomization. Inspired by that work, we propose a discrete-phase-randomized (DPR) MDI-QKD protocol for solving the phase randomization loophole in MDI-QKD. In addition, we provide a formal security proof of the DPR MDI-QKD protocol.

The roadmap for the rest of the paper is as follows. In Sec.~\ref{sec:reviewMDIQKD}, we first provide a brief review of the MDI-QKD protocol. In Sec.~\ref{sec:attack}, we show an attack against a MDI-QKD system with imperfect phase randomization. In Sec.~\ref{sec:protocol}, we describe the DPR MDI-QKD protocol and provide its security analysis. In Sec.~\ref{sec:conclusion}, we summarize the results and discuss future work.

\section{Review of MDI-QKD}
\label{sec:reviewMDIQKD}
\begin{figure}[htb]
\centering \includegraphics[width=7cm]{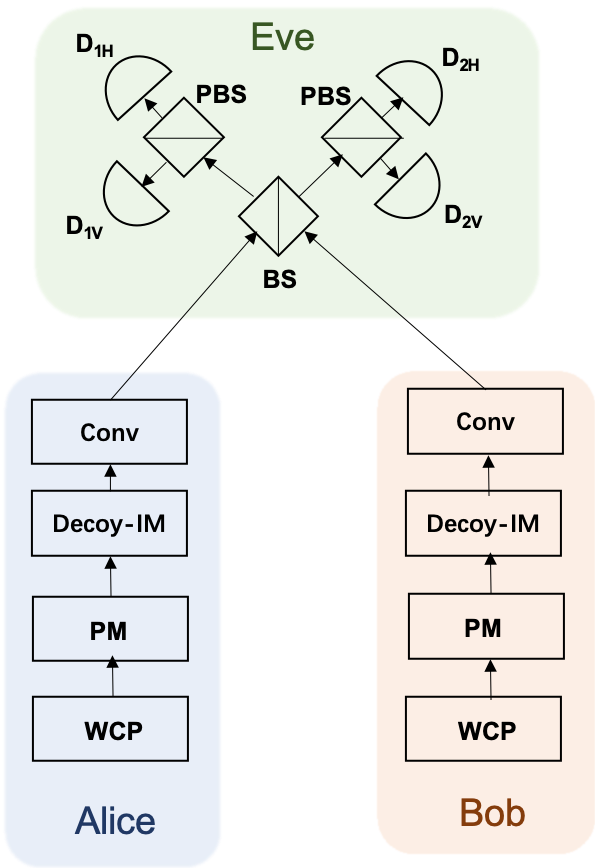}
\caption{Diagram of the MDI-QKD protocol. There are three modules, including Alice, Bob and Eve. Alice and Bob generate source states, whereas the measurement device controlled by Eve performs a joint measurement on the states of Alice and Bob. Here, WCP stands for a weak coherent source; PM stands for a phase modulator; Decoy-IM stands for an intensity modulator that switches between the signal states and the decoy states. Conv is a converter that converts the phase encoding to the polarization encoding \cite{tamaki2012phase}; BS is a beam splitter; PBS is a polarization beam splitter; $D_{1H}$, $D_{1V}$, $D_{2H}$, $D_{2V}$ are single-photon detectors. } 
\label{fig:mdiqkd}
\end{figure}

A diagram of the MDI-QKD protocol is shown in Fig.~\ref{fig:mdiqkd}. In a typical MDI-QKD setup, Alice and Bob prepare source states in the rectilinear basis or  in the diagonal basis. A measurement device which may be controlled by Eve performs a joint Bell measurement on Alice's and Bob's states, and outputs either $\ket{01}+\ket{10}$ or $\ket{01} -\ket{10}$ (other Bell measurement outcomes are discarded). Afterwards, Alice and Bob announce the bases they used and discard the events that they use different bases. It can be shown that if both Alice and Bob were using the rectilinear basis with different (the same) eigenstates, the measurement result is always $\ket{01} -\ket{10}$ ($\ket{01}+\ket{10}$). If both parties were using  the diagonal bases, then outputting $\ket{01}+\ket{10}$ and $\ket{01} -\ket{10}$ will have the same probability  0.5. By these properties, Alice and Bob can use the rectilinear basis to generate keys, and use the diagonal basis to estimate the errors in the measurement device. In addition, the two parties use the decoy state method \cite{Hwang:Decoy:2003,Lo:Decoy:2005,Wang:Decoy:2005} to estimate the channel gain and error rate with higher precision.  The MDI-QKD protocol can be viewed as a time-reversed version of an entanglement-based QKD protocol \cite{Ekert:QKD:1991} and indeed its security can be proved using this time-reversal symmetry~\cite{Lo:MDIQKD:2012}.

\section{Vulnerability of imperfect phase randomization}
\label{sec:attack}
In this section, we propose an attack to show that there is a serious loophole in MDI-QKD if the phase of the coherent source is not properly randomized.
For simplicity, we consider the extreme case that there is no phase randomization, and the phases of the signal state and the decoy state are known to the eavesdropper Eve. We now describe how to use unambiguous state discrimination (USD) to attack a MDI-QKD system without phase randomization. 

In the first step, Eve uses USD to distinguish the signal state and the decoy state on both Alice's and Bob's sides, each with some probability $q$ (note that in the case of perfect phase randomization, Eve cannot distinguish the signal state and the decoy state, i.e., $q=0$). Eve discards the events that he fails to distinguish the signal state and the decoy state. Then Eve measures the photon number and chooses to forward some of the photons conditioned on the results of signal or decoy states and the photon number to preserve the channel statistics. 

In a normal MDI-QKD, the key rate is lower bounded by 
\begin{equation}
\label{Discrete:originalMDIQKD}
R^l = -Q_{\textrm{Rect}} f(E_{\textrm{Rect}}) H(E_{\textrm{Rect}}) + Q^{1,1}_n( 1 - H(E^{1,1,b}_{\textrm{Diag}}) ),
\end{equation}
where the first term is  the error correction term and the second term is the privacy amplification term.
Here $Q_{\textrm{Rect}}$ is the gain of the rectilinear basis, $E_{\textrm{Rect}}$ is the bit error rate of the rectilinear basis, $Q^{1,1}_n$ is the estimated gain when both parities emit single-photon states, $f(\cdot) \ge 1$ is the error correction efficiency, $E^{1,1,b}_{\textrm{Diag}}$ is the bit error rate of the diagonal basis, and $H(\cdot)$ is the binary Shannon entropy.
Under the attack, the key rate is upper bounded by $R^u= Q^{1,1}_a$ where $Q^{1,1}_a$ is the actual gain of the single photon states from both parities under the attack. Apparently, if $R_l > R^u$, then Alice and Bob would mistakenly generate keys with a key rate higher than the maximal secure key rate possible under the attack, thus leaking part of the key information to Eve. 
The goal of Eve is hence to minimize $R^u$ to the extent that it is smaller than $R^l$.
We next show that this indeed can happen.

Suppose the  intensities of the signal state and the decoy state are $\mu$ and $\nu$ respectively, it can be shown~\cite {tang2013source} that on each side, the optimal success probability of unambiguous state discrimination is 
\begin{equation}
q_{\textrm{opt}} = 1 - \textrm{exp}(-\frac{|\sqrt{\mu} - \sqrt{\nu} |^2}{4}).
\end{equation}
In the attack, the gains of the signal state and the decoy state at each side are 
\begin{eqnarray} \label{Discrete:USDattack}
Q_\mu &= &  \sum\limits_{i=1}^{\infty} q_{opt} Z_i^\mu e^{-\mu} \frac{\mu^i}{i!},  \nonumber \\
Q_\nu &=&  \sum\limits_{i=1}^{\infty} q_{opt} Z_i^\nu e^{-\nu} \frac{\nu^i}{i!} 
\end{eqnarray}
respectively, where $Z_i^\mu$ ($Z_i^\nu$) is the probability of Eve forwarding the photons conditioned on the signal state (the decoy state) and the photon number $i$. Here we make the simplified assumption that the dark count is zero, so the summation index starts from 1.

Eve should choose $Z_i^{\mu}$ and $Z_i^{\nu}$ properly so that his faked gains match the normal channel gains of both the signal state and the decoy state, namely,
\begin{eqnarray} \label{Discrete:USDnormal}
Q_\mu &= &  1 - e^{-\eta \mu},  \nonumber \\
Q_\nu &=&  1 - e^{-\eta \nu}.
\end{eqnarray}
Here $\eta$ is the channel loss and we assume there is no dark count for simplicity.
In addition, since 
\begin{equation}
R^u = Q^{1,1}_a = (  q_\mu Z_1^\mu e^{-\mu}\mu)^2,
\end{equation} 
minimizing $R^u$ is equivalent to minimizing $Z_1^\mu$. 

Assume $\mu \ll 1$ and $\mu > \nu >  \mu^2 /2 $ and let $\eta = q_{\textrm{opt}} \mu /2$, we can take 
\begin{eqnarray} 
Z_2^{\mu} &= &  1,  \nonumber \\
Z_1^{\nu} &= &  \mu^2/2\nu,  \nonumber \\
Z_i^{\mu} &= &  0 \;  \forall i \neq 2, \\
Z_i^{\nu} &= &  0 \; \forall i \neq 1.  \nonumber 
\end{eqnarray}
For these parameters, it can be checked that the constraints Eqs.~\eqref{Discrete:USDattack} to \eqref{Discrete:USDnormal} are satisfied.
Hence we have $Z_1^\mu=0$, thus $R^u = 0$, meaning that all the key information is leaked to Eve. It only remains to show that $R^l > 0$ for these parameters.

For simplicity, we assume there are no errors, namely $E_\mu = 0$. The estimated key rate lower bound $R^l$ is then reduced to $Q^{1,1}_n$. In a normal estimation, 
since Eve cannot distinguish the signal state and the decoy state, we have
\begin{eqnarray} \label{Discrete:normalDist}
Q_{\mu, \mu} &= &  \sum\limits_{i.j=1}^{\infty} Y_{i,j} e^{-\mu} \frac{\mu^i}{i!} e^{-\mu} \frac{\mu^j}{j!},  \nonumber \\
Q_{\nu,\mu}  &=&  \sum\limits_{i,j=1}^{\infty} Y_{i,j} e^{-\nu} \frac{\nu^i}{i!}e^{-\mu} \frac{\mu^j}{j!},   \nonumber \\
 Q_{\mu,\nu}  &=&  \sum\limits_{i,j=1}^{\infty} Y_{i,j} e^{-\mu} \frac{\mu^i}{i!}e^{-\nu} \frac{\nu^j}{j!},  \\
Q_{\mu, \mu} &= &  \sum\limits_{i.j=1}^{\infty} Y_{i,j} e^{-\mu} \frac{\mu^i}{i!} e^{-\mu} \frac{\mu^j}{j!}.  \nonumber 
\end{eqnarray}
Here $Q_{\alpha,\beta}$ stands for the gain when the mean photon number of Alice's state is $\alpha$ and the mean photon number  of Bob's state is $\beta$, $Y_{i,j}$ stands for the gain when Alice's state contains $i$ photon and Bob's state contains $j$ photons. Since Alice and Bob send their states independently, we have $Q_{\alpha,\beta} = Q_\alpha Q_\beta$,
where $Q_\alpha$ and $Q_\beta$ are given by Eq.~\eqref{Discrete:USDnormal}.

By a two-step estimation, we first estimate the intermediate quantities $Y^1_\mu$ and $Y^1_\nu$ defined by
\begin{eqnarray} \label{Discrete:Intermediate}
Y^1_\mu &= &  \sum\limits_{i=1}^{\infty} Y_{1,i} e^{-\mu} \frac{\mu^i}{i!},  \nonumber \\
Y^1_\nu &=&  \sum\limits_{i=1}^{\infty} Y_{1,i} e^{-\nu} \frac{\nu^i}{i!}.  
\end{eqnarray}
Using Eq.~\eqref{Discrete:normalDist},  $Y^1_\mu$ can be estimated from $Q_{\mu,\mu}$ and $Q_{\nu,\mu}$ as
\begin{eqnarray} 
\label{Discrete:Y1mu}
Y^1_\mu & \ge & \frac{\mu}{ \mu\nu -\nu ^2 } (Q_{\nu,\mu} e^{\nu} - Q_{\mu,\mu} e^\mu \frac{\nu^2}{\mu^2}) \nonumber \\
&  \approx  &   \eta^2 \mu,  
\end{eqnarray}
and similarly   for $Y^1_\nu$, we have 
\begin{equation}
\label{Discrete:Y1nu}
Y^1_\nu  \approx \eta^2 \nu.
\end{equation} 
Finally, by Eqs.~\eqref{Discrete:Intermediate} to \eqref{Discrete:Y1nu}, $Y_{1,1}$ can be estimated as
\begin{eqnarray} 
Y_{1,1} & \ge & \frac{\mu}{ \mu\nu -\nu ^2 } (Y^1_\nu e^{\nu} - Y^1_\mu e^\mu \frac{\nu^2}{\mu^2}) \nonumber \\
&  \approx  &   \eta^2.  
\end{eqnarray}
Thus $R^l = Q_{1,1} = Y_{1,1} (e^{-\mu} \mu )^2 > 0 = R^u$, which shows that 
Eve's attack is successful. 

It should be noted that this example is not the only case that Eve can successfully attack a MDI-QKD system without phase randomization. The exact parameter region which is vulnerable to Eve's attack is beyond the scope of this paper, and is left as an interesting future research direction.

\section{Discrete-phase-randomized MDI-QKD protocol}
\label{sec:protocol}
In this section, we first describe our discrete-phase-randomized MDI-QKD protocol and then provide its security analysis.

A weak coherent laser can be described by the following state~\cite {PhysRev.131.2766}
\begin{equation} \label{Discrete:CoherentState}
\ket{\alpha}=e^{-\frac{|\alpha|^2}{2}}\sum \frac{\alpha^n}{\sqrt{n!}}\ket{n},
\end{equation}
where $\alpha$ is a complex number and $\ket{n}$ is the Fock state of $n$ photons. 
In continuous phase randomization, a random phase $\theta\in [0,2\pi)$ is applied on $\ket{\alpha}$, and the input state 
becomes
\begin{equation} \label{Discrete:ContinuousPhase}
\frac{1}{2 \pi}\int_{0}^{2\pi}\ket{\alpha e^{i\theta}}\bra{\alpha e^{i\theta}}d\theta =\sum\limits_{n=0}^{\infty}e^{-|\alpha|^2}\frac{|\alpha|^2}{n!}\ket{n}\bra{n}.
\end{equation}
Conditioning on sufficiently small $|\alpha|$ and photon detection, this input state approximates the single photon state $\ket{1}\bra{1}$ quite well and hence is a good substitute for a single photon source.

In contrast to continuous phase randomization, in our discrete-phase-randomized MDI-QKD protocol, we apply one of the discrete phases
\begin{equation} \label{Discrete:Nphases}
\{\theta_k=\frac{2\pi k}{N} | k=0,1,\dots,N-1\}
\end{equation}
randomly on the weak coherent laser $\ket{\sqrt{2}\alpha}$. Here $N$ is the number of discrete phases. Using the virtual qudit formalism of randomization,
the input state can be written as 
\begin{eqnarray} \label{Discrete:Nrho}
\ket{\Psi_N} &=& \sum\limits_{k=0}^{N-1}\ket{a_k}_A\ket{\sqrt{2}\alpha e^{2k\pi i/N}}_B \\
&=& \sum\limits_{j=0}^{N-1} \ket{b_j}_A \ket{\lambda_j}_B, \nonumber
\end{eqnarray}
where
\begin{equation} \label{Discrete:modNrho}
\ket{\lambda_j} = \sum_{k=0}^{N-1} e^{-2kj\pi i/N}\ket{e^{2k\pi i/N}\sqrt{2}\alpha}. \\
\end{equation}
Here $\{\ket{a_k}\}_{k = 0,1,\dots,N-1}$ and $\{\ket{b_j}\}_{j = 0,1,\dots,N-1}$ are sets of orthogonal bases,
and $\ket{a_k}$ can be transformed from $\ket{b_j}$ by
\begin{equation}
\ket{a_k} = \sum_{j=0}^{N-1} e^{-2kj\pi i/N} \ket{b_j}.
\end{equation}

By Taylor expansion on $\ket{\lambda_j}$, one has
\begin{equation} \label{Discrete:rhojFock}
\ket{\lambda_j}=\sum_{l=0}^{\infty} \frac{(\sqrt{2}\alpha)^{lN+j}}{\sqrt{(lN+j)!}}\ket{lN+j}. \\
\end{equation}
The probability of obtaining $\ket{\lambda_j}$ is
\begin{eqnarray} \label{Discrete:Probj}
P_j = \frac{\braket{\lambda_j}{\lambda_j}}{\sum_{j=0}^{N-1} \braket{\lambda_j}{\lambda_j}} 
= \sum_{l=0}^{\infty}\frac{\mu^{lN+j}e^{-\mu}}{(lN+j)!},
\end{eqnarray}
where $\mu = 2|\alpha|^2$. It can be seen that as $N$ goes to infinity, $\ket{\lambda_j}$ approaches the Fock state $\ket{j}$.  Therefore we will call $\ket{\lambda_j}$  the approximated $j$-photon state.

The input state $\ket{\lambda_j}$ is then encoded into four BB84 states with the phase encoding and becomes one of
\begin{eqnarray} \label{Discrete:BB84logic4}
\ket{0_x^L} &=& \sum_{k=0}^{N-1} e^{-2kj\pi i/N}\ket{e^{2k\pi i/N}\alpha} \ket{e^{2k\pi i/N}\alpha}, \nonumber \\
\ket{1_x^L} &= &\sum_{k=0}^{N-1} e^{-2kj\pi i/N}\ket{e^{2k\pi i/N}\alpha} \ket{-e^{2k\pi i/N}\alpha}, \nonumber \\
\ket{0_y^L} &= &\sum_{k=0}^{N-1} e^{-2kj\pi i/N}\ket{e^{2k\pi i/N}\alpha} \ket{ie^{2k\pi i/N}\alpha}, \\
\ket{1_y^L} &= &\sum_{k=0}^{N-1} e^{-2kj\pi i/N}\ket{e^{2k\pi i/N}\alpha} \ket{-ie^{2k\pi i/N}\alpha},\nonumber
\end{eqnarray}
where $\ket{0_x^L}$ and $\ket{1_x^L}$ are logical qubits in the $X$ basis, and $\ket{0_y^L}$ and $\ket{1_y^L}$ are logical qubits in the $Y$ basis, the first coherent state is the reference state and the second coherent state is the signal state with BB84 phases. Since the probabilities of choosing the eigenstates are equal, the overall states encoded in the $X$ basis and the $Y$ basis are
\begin{eqnarray} \label{Discrete:rhoxy}
\rho_{AB}^X &= & (\ket{0_x^L}\bra{0_x^L}+\ket{1_x^L}\bra{1_x^L})_A\otimes(\ket{0_x^L}\bra{0_x^L}+\ket{1_x^L}\bra{1_x^L})_B,  \nonumber \\
\rho_{AB}^Y &=& (\ket{0_y^L}\bra{0_y^L}+\ket{1_y^L}\bra{1_y^L})_A \otimes(\ket{0_y^L}\bra{0_y^L}+\ket{1_y^L}\bra{1_y^L})_B,  \nonumber \\
\end{eqnarray}
respectively.
In the ideal case of basis-independent sources, we have
\begin{equation} \label{Discrete:BasisInd}
\rho_{AB}^X = \rho_{AB}^Y.
\end{equation}
We can characterize the deviation from the ideal case by bounding the fidelity between $\rho_{AB}^X$ and $\rho_{AB}^Y$ as
\begin{eqnarray}
\label{eq:fidelity}
&F_{j,j}(\rho_{AB}^X, \rho_{AB}^Y) =\mathrm{tr} \sqrt{\sqrt{\rho_{AB}^Y} \rho_{AB}^X \sqrt{\rho_{AB}^Y}} \label{Discrete:FidelityXY}\\
&\ge \left| \frac{\sum_{l=0}^{\infty} \frac{\mu^{lN+j}}{(lN+j)!} 2^{-\frac{lN+j}{2}} \left(\cos\frac{lN+j}{4}\pi+\sin\frac{lN+j}{4}\pi\right) }{\sum_{l=0}^{\infty}\frac{\mu^{lN+j}}{(lN+j)!}} \right|^2. \nonumber
\end{eqnarray}
The concrete derivation can be found in Appendix~\ref{AppSec:Formula}.
The first order approximation of $F_{1,1}$ with respect to $\mu^N$ is 
\begin{equation} \label{Discrete:F1st01}
F^{(1)}_{1,1} \ge 1 -2\left(1-2^{-\frac{N}{2}} \cos\frac{N}{4}\pi \right)\frac{\mu^N}{(N+1)!} .
\end{equation}
This will be later used in the key rate formula. Its derivation can also be found in Appendix~\ref{AppSec:Formula}.

\subsection{Key rate}   
In a normal MDI-QKD, the key rate formula is given by Eq.~\eqref{Discrete:originalMDIQKD}.
In the discrete phase version, we need to modify the key rate formula to
\begin{eqnarray} \label{Discrete:GLLPmodif}
R& \ge & -  Q_{\textrm{Rect}} f(E_{\textrm{Rect}}) H(E_{\textrm{Rect}}) \nonumber \\
 &  & +   \sum_i \sum_j P_i P_j Y_{i,j} [1-H(E^{i,j,p}_{\textrm{Rect}})].
\end{eqnarray}
The error correction part stays unchanged. For the privacy amplification part, 
$P_i$ is the probability of obtaining the state $\ket{\lambda_i}$ when a party uses a signal state, $Y_{i,j}$ and $E^{i,j,p}_{\textrm{Rect}}$ are the gain and the phase error rate of the rectilinear basis when Alice's state is $\ket{\lambda_i}$, Bob's state is $\ket{\lambda_j}$ and both parties use signal states.

Recall that a phase error of the rectilinear basis occurs when Alice and Bob's states are both encoded in the $X$ basis, and their joint state after the Bell measurement is $\ket{0_x^L}_A\ket{0_x^L}_B+\ket{1_x^L}_A\ket{1_x^L}_B$ instead of the correct outcome $\ket{0_x^L}_A\ket{0_x^L}_B-\ket{1_x^L}_A\ket{1_x^L}_B$. If their joint state after the Bell measurement is $\ket{0_x^L}_A\ket{1_x^L}_B-\ket{0_x^L}_A\ket{1_x^L}_B$, a bit error of the rectilinear basis is said to occur. Similarly, for the diagonal basis where Alice and Bob's states are both encoded in the $Y$ basis, the correct outcome after the Bell measurement should be $\ket{0_y^L}_A\ket{0_y^L}_B-\ket{1_y^L}_A\ket{1_y^L}_B$. If the actual joint state is $\ket{0_y^L}_A\ket{0_y^L}_B+\ket{1_y^L}_A\ket{1_y^L}_B$, a phase error of the diagonal basis is said to occur. If the actual joint state is $\ket{0_y^L}_A\ket{1_y^L}_B-\ket{1_y^L}_A\ket{0_y^L}_B$, a bit error of the diagonal basis is said to occur.

In the key formula, since $Q_{\textrm{Rect}}$ and $E_{\textrm{Rect}}$ can be directly measured, only $Y_{i,j}$ and $E^{i,j,p}_{\textrm{Rect}}$ need to be estimated. In the basis-independent case, $E^{1,1,p}_{\textrm{Rect}} = E^{1,1,b}_{\textrm{Diag}}$, hence the phase error of the rectilinear basis can be estimated using the bit error rate of the diagonal basis. However, in the discrete phase case, the basis independence  property no longer holds. Fortunately,  we can estimate the difference between $e_{1,1}^b=E^{1,1,b}_{\textrm{Diag}}$ and $e_{1,1}^p = E^{1,1,p}_{\textrm{Rect}}$ as follows~\cite{LoPreskill:NonRan:2007},
\begin{eqnarray} \label{Discrete:bitphase}
e_{j,j}^p \le & e_{j,j}^b+4\Delta_{j,j}(1-\Delta_{j,j})(1-2e_{j,j}^b)   \nonumber\\
  & +4(1-2\Delta_{j,j})\sqrt{\Delta_{j,j}(1-\Delta_{j,j}) e_{j,j}^b(1-e_{j,j}^b)},
\end{eqnarray}
where 
\begin{equation} \label{Discrete:bias}
\Delta_{j,j} = \frac{1-F_{j,j}}{2Y_{j,j}}.
\end{equation}
 Here $F_{j,j}$ is given by Eq.~\eqref{eq:fidelity}. 
Next we show how to estimate the parameters $Y_{1,1}$ and $e_{1,1}^b$.

\subsection{ Parameter estimation}

In discrete-phase-randomized MDI-QKD, we need to estimate the gain $Y^{\alpha, \beta}_{i,j}$ and the error rate $e^{\alpha, \beta}_{i,j}$.
First we note that the following relations hold: 
\begin{eqnarray} \label{Discrete:QESD}
Q^{\alpha, \beta} &=& \sum_{i,j=0}^{N-1} P^\alpha_{i} P^\beta_{j} Y^{\alpha, \beta}_{i,j}, \nonumber \\
Q^{\alpha, \beta} E^{\alpha, \beta}&=& \sum_{i,j=0}^{N-1}P^\alpha_{i} P^\beta_{j} Y^{\alpha, \beta}_{i,j}e^{\alpha, \beta}_{i,j},
\end{eqnarray}
where $\alpha$($\beta$) distinguishes the signal state and the decoy states, $i$($j$) stands for the approximated $i$-photon($j$-photon) state, $Q^{\alpha, \beta}$ and $E^{\alpha, \beta}$ are the observed gain and error rate in the case that Alice uses the intensity setting $\alpha$ and Bob uses the intensity setting $\beta$, $Y^{\alpha, \beta}_{i,j}$ and $e^{\alpha, \beta}_{i,j}$ are the intrinsic gain and error rate in the case that Alice uses the intensity setting $\alpha$ and the approximated $i$-photon state, Bob uses the intensity setting $ \beta$ and  the approximated $j$-photon state, $P^\alpha_{i}$ is the probability of generating an approximated $i$-photon state when the intensity setting is $\alpha$.
 
There is an inherent assumption in normal MDI-QKD, namely
\begin{eqnarray} \label{Discrete:DecoyAssum}
Y_{i,j}^{\alpha_1, \beta_1} &= Y_{i,j}^{\alpha_2, \beta_2}, \nonumber \\
e_{i,j}^{\alpha_1, \beta_1}&= e_{i,j}^{\alpha_2, \beta_2}.
\end{eqnarray}
This no longer holds in the case of discrete phase randomization as 
\begin{equation} \label{Discrete:AssumFail}
\ket{\lambda_{i,j}^{\alpha_1, \beta_1}} \neq \ket{\lambda_{i,j}^{\alpha_2, \beta_2}},
\end{equation}
where $\ket{\lambda_{i,j}^{\alpha, \beta}}$ is the joint state of Alice and Bob when Alice uses the intensity setting $\alpha$ together with the approximated $i$-photon state, and Bob uses the intensity setting $\beta$ together with the approximated $j$-photon state.
Nevertheless, we can bound the difference between gains and errors of different intensities as
\begin{eqnarray} \label{Discrete:DecoyBound}
|Y_{i,j}^{\alpha,\mu}-Y_{i,j}^{\alpha,\nu}|   & \le   \sqrt{1- F^{2}_{\mu\nu}} , \nonumber \\
|Y_{i,j}^{\alpha,\mu}e_{i,j}^{\alpha,\mu}-Y_{i,j}^{\alpha,\nu}e_{i,j}^{\alpha,\nu}|   & \le  \sqrt{1- F^{2}_{\mu\nu}}   , 
\end{eqnarray}
where
\begin{equation} \label{Discrete:Fmunu}
F_{\mu\nu} = \frac{\sum_{l=0}^{\infty} \frac{(\mu\nu)^{lN/2}}{(lN)!}}{\sqrt{\sum_{l=0}^{\infty} \frac{\mu^{lN}}{(lN)!}\sum_{l=0}^{\infty} \frac{\nu^{lN}}{(lN)!}}}.
\end{equation}
The derivation of these bounds can be found in Appendix \ref{AppSec:DecoyBound}.

The estimation of the gain $Y^{\alpha, \beta}_{i,j}$ and the error rate $e^{\alpha, \beta}_{i,j}$ is similar to normal MDI-QKD. We start with the estimation of the gain $Y^{\alpha, \beta}_{i,j}$. Note that the first equation in Eq.~\eqref{Discrete:QESD} can be rewritten as
\begin{equation}
\label{Discrete:firstlevelgain}
Q^{\alpha, \beta} = \sum_{i=0}^{N-1} P^\alpha_{i} Y^{\alpha, \beta}_{i},
\end{equation}
where
\begin{equation}
\label{Discrete:secondlevelgain}
Y^{\alpha, \beta}_i = \sum_{j=0}^{N-1} P^\beta_{j} Y^{\alpha, \beta}_{i,j}.
\end{equation}
For notation simplicity, let
\begin{equation}
\epsilon= \sqrt{1- F^{2}_{\mu\nu}}.
\end{equation}
From Eq.~\eqref{Discrete:secondlevelgain}, we get
\begin{eqnarray}
|Y^{\alpha, \beta}_i - Y^{\mu, \beta}_i | & = & | \sum_{j=0}^{N-1} P^\beta_{j} (Y^{\alpha, \beta}_{i,j}-Y^{\mu, \beta}_{i,j}) | \nonumber \\
& \le & \sum_{j=0}^{N-1} P^\beta_{j}|Y^{\alpha, \beta}_{i,j}-Y^{\mu, \beta}_{i,j} | \\
& \le &  \sum_{j=0}^{N-1} P^\beta_{j} \epsilon = \epsilon, \nonumber
\end{eqnarray}
where the last inequality holds because $ \sum_{j=0}^{N-1} P_j^\beta = 1$.
Hence, we can estimate the upper bound and the lower bound of $Y^{\mu, \beta}_i $ under the following constraint,
\begin{eqnarray}
Q^{\mu, \beta} &=& \sum_{i=0}^{N-1} P^\mu_{i} Y^{\mu, \beta}_{i}, \nonumber \\
Q^{\alpha, \beta} &=& \sum_{i=0}^{N-1} P^\alpha_{i} Y^{\alpha, \beta}_{i} =\sum_{i=0}^{N-1} P^\alpha_{i} Y^{\mu, \beta}_{i} \pm \epsilon, \\
 0&\le &Y^{\mu, \beta}_{i} \le 1. \nonumber
\end{eqnarray}
After the range of $Y^{\mu, \beta}_i $ is estimated for all $\beta$, we can estimate the upper bound and the lower bound of $Y^{\mu, \mu}_{i,j}$
under the following constraint:
\begin{eqnarray}
Y^{\mu, \mu}_i &=& \sum_{j=0}^{N-1} P^\mu_{j} Y^{\mu, \mu}_{i,j}, \nonumber \\
Y^{\mu, \beta}_i &=& \sum_{j=0}^{N-1} P^\beta_{j} Y^{\mu, \beta}_{i,j}= \sum_{j=0}^{N-1} P^\beta_{j} Y^{\mu, \mu}_{i,j} \pm \epsilon, \\
 0&\le &Y^{\mu, \mu}_{i,j} \le 1. \nonumber
\end{eqnarray}

The estimation of $Y^{\alpha, \beta}_{i,j}e^{\alpha, \beta}_{i,j}$ is almost identical to the estimation of $Y^{\alpha, \beta}_{i,j}$. 
We can rewrite the second equation in Eq.~\eqref{Discrete:QESD} as 
\begin{equation}
\label{Discrete:firstlevelerror}
Q^{\alpha, \beta} E^{\alpha, \beta}= \sum_{i}^{N-1}P^\alpha_{i}  W^{\alpha, \beta}_{i}  ,
\end{equation}
where
\begin{equation}
\label{Discrete:secondlevelerror}
W^{\alpha, \beta}_{i} = \sum_{j=0}^{N-1}P^\beta_{j} Y^{\alpha, \beta}_{i,j}e^{\alpha, \beta}_{i,j}.
\end{equation}
From Eq.~\eqref{Discrete:secondlevelerror}, we get
\begin{eqnarray}
|W^{\alpha, \beta}_i - W^{\mu, \beta}_i | & = & | \sum_{j=0}^{N-1} P^\beta_{j} (Y^{\alpha, \beta}_{i,j}e^{\alpha, \beta}_{i,j}-Y^{\mu, \beta}_{i,j}e^{\mu, \beta}_{i,j}) | \nonumber \\
& \le & \sum_{j=0}^{N-1} P^\beta_{j}|Y^{\alpha, \beta}_{i,j}e^{\alpha, \beta}_{i,j}-Y^{\mu, \beta}_{i,j}e^{\mu, \beta}_{i,j} | \\
& \le &  \sum_{j=0}^{N-1} P^\beta_{j} \epsilon = \epsilon. \nonumber
\end{eqnarray}
Hence, we can estimate  the upper bound and the lower bound of $W^{\mu, \beta}_i $ under the following constraint:
\begin{eqnarray}
Q^{\mu, \beta} E^{\mu, \beta} &=& \sum_{i=0}^{N-1} P^\mu_{i} W^{\mu, \beta}_{i}, \nonumber \\
Q^{\alpha, \beta} E^{\alpha, \beta}&=& \sum_{i=0}^{N-1} P^\alpha_{i} W^{\alpha, \beta}_{i} =\sum_{i=0}^{N-1} P^\alpha_{i} W^{\mu, \beta}_{i} \pm \epsilon, \\
 0&\le &W^{\mu, \beta}_{i} \le 1. \nonumber
\end{eqnarray}
After the range of $W^{\mu, \beta}_i $ is estimated for all $\beta$, we can estimate the upper bound and the lower bound of $Y^{\mu, \mu}_{i,j}e^{\mu, \mu}_{i,j}$
under the following constraint:
\begin{eqnarray}
W^{\mu, \mu}_i &=& \sum_{j=0}^{N-1} P^\mu_{j} Y^{\mu, \mu}_{i,j}e^{\mu, \mu}_{i,j}, \nonumber \\
W^{\mu, \beta}_i &=& \sum_{j=0}^{N-1} P^\beta_{j} Y^{\mu, \beta}_{i,j}e^{\mu, \beta}_{i,j}= \sum_{j=0}^{N-1} P^\beta_{j} Y^{\mu, \mu}_{i,j} e^{\mu, \mu}_{i,j}\pm \epsilon,  \label{eq:finalestimate} \\
 0&\le &Y^{\mu, \mu}_{i,j}e^{\mu, \mu}_{i,j} \le 1. \nonumber
\end{eqnarray}
This completes the parameter estimation of the discrete-phase-randomized MDI-QKD protocol.

Each linear system presented in this section can be efficiently solved through linear programming. When there are $M$ decoy states, each linear system contains $N$ variables and $2N+2M+1$ constraints. Hence, the computation of its solution is manageable when $N$ is small (e.g., $N<20$). When $N$ is large (e.g., $N>10\;000$), the computation time can be infeasibly long. In that case, one method to accelerate the computation at a cost of a small decrease in performance is that we keep only variables with the lowest $K$ indices, such as $Y_0^{\mu,\beta},\dots, Y_{K-1}^{\mu,\beta}$, and relax other variables to 0 or 1 in all constraining equations and inequalities. The reduced linear system then contains $K$ variables and $2K+2M+2$ constraints. In later simulations, we take $M=2$ and $K=3$. Larger values of $M$ and $K$ can lead to more accurate estimation of the parameters.

\subsection{Simulation result}
\begin{figure}[htb]
\centering \includegraphics[width=9cm]{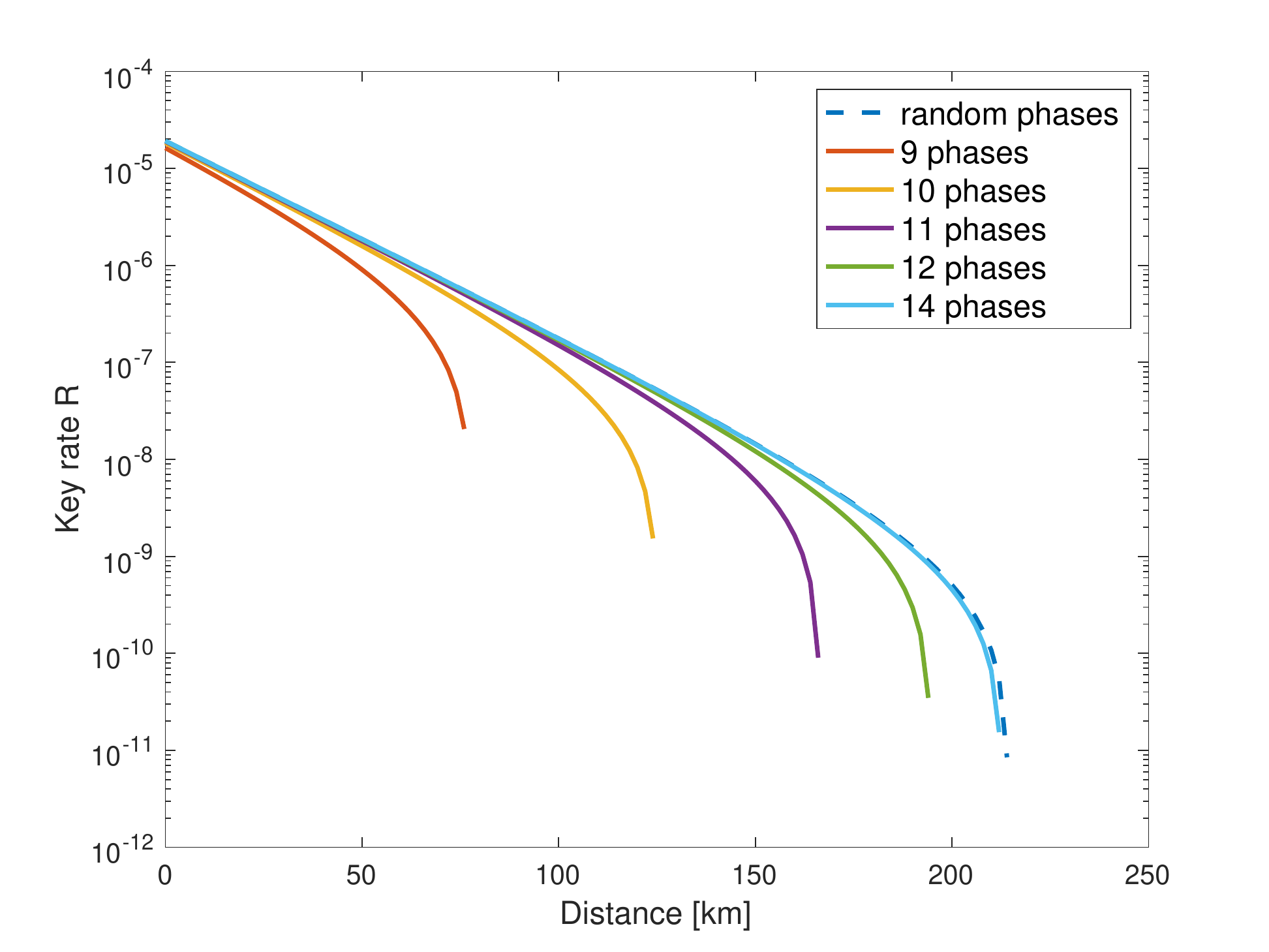}
\caption{The relation between the key rate and the transmission distance for continuously random phases and discrete phases. The dashed line is the key rate for continuously random phases and the solid lines from left to right are for 9, 10, 11, 12, and 14 discrete phases, respectively. } 
\label{Fig:ExpScheme}
\end{figure}

In Fig.~\ref{Fig:ExpScheme}, we plot the key rate of continuous randomization (annotated as ``random phases'') and various number of discrete phases (9, 10, 11, 12, 14 phases, respectively) under different transmission distances. It can be seen that 14 phases already approximate continuous phase randomization quite well. The detailed simulation model and simulation parameters are shown in  Appendix \ref{AppSec:Sim}. 

\begin{figure}[htb]
\centering \includegraphics[width=9cm]{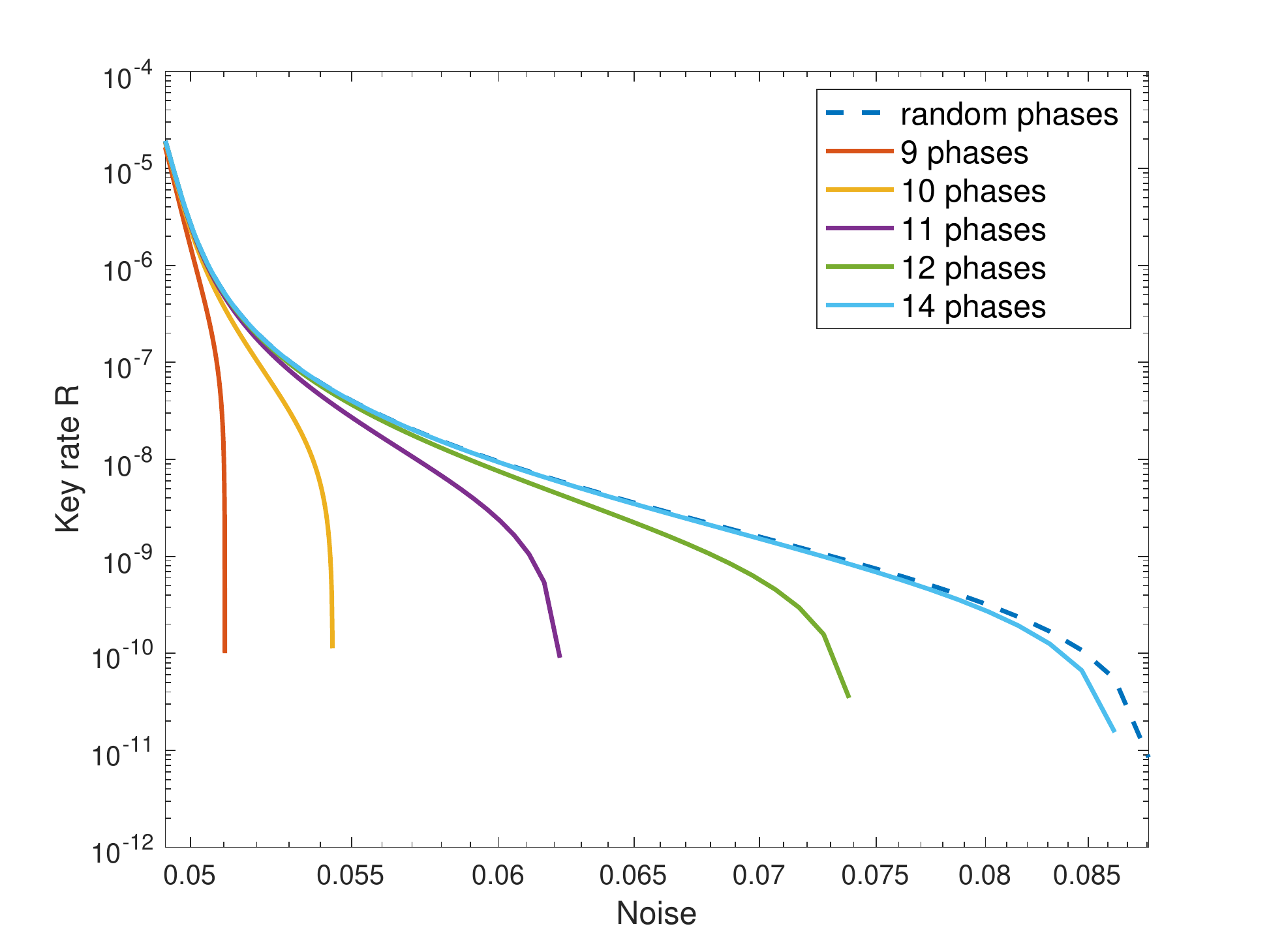}
\caption{The relation between the key rate and the noise level for continuously random phases and discrete phases. The dashed line is the key rate for continuously random phases and the solid lines from left to right are for 9, 10, 11, 12, and 14 discrete phases, respectively. } 
\label{Fig:ExpScheme2}
\end{figure}

With the same simulation model, we plot the key rate of continuous randomization (annotated as ``random phases'') and various number of discrete phases (9, 10, 11, 12, 14 phases, respectively) under different noise levels $e_{1,1}^b$ in Fig.~\ref{Fig:ExpScheme2}. It can be seen that the security threshold (maximally tolerable noise) of 14 phases is already very close to that of continuous phase randomization, which is about 8.7\%.

In a practical experiment, the deviation of experimental parameters from the simulation parameters used here should be accounted for by substituting the actual experimental parameters into the simulation model, and the selection of the number of discrete phases should be determined through this revized simulation.

\section{Conclusion}
\label{sec:conclusion}
In summary, we showed that MDI-QKD with imperfect phase randomization is vulnerable to attacks and, as a solution, proposed a discrete-phase-randomized measurement-device-independent quantum key distribution protocol. 
We also provided a security proof of the protocol. Simulation results confirm that the protocol with only a few phases (14 phases) already approximates continuous phase randomization quite well. 

As future work, we can consider further source imperfection in measurement-device-independent quantum key distribution. One direction is to consider imperfectly prepared discrete phases $\{ \ket{\alpha e^{i(2\pi k/N\pm \delta)}}\}_{k=1,\dots,N}$, where $\delta$ is a small quantity characterizing the deviation from the exact discrete phases. One can modify the fidelity calculation to accommodate for this change.  Another direction is to extend our analysis to other MDI protocols requiring weak coherent sources, such as MDI entanglement witness \cite{branciard2013measurement}.

\section*{Acknowledgements}
This work was supported by the internal Grant No. SLH00202007 from East China University of Science and Technology.

\appendix

\section{Fidelity Calculation}
\label{AppSec:Formula}

In this section, we will provide the details on the calculation of the fidelity between the input states prepared in different bases.
We will utilize a few results from Ref.~\cite{cao2015discrete}.

By Eqs.~\eqref{Discrete:rhoxy} and \eqref{eq:fidelity} in the main text, we have
\begin{eqnarray}
&&F_{j,j}(\rho_{AB}^X, \rho_{AB}^Y)  \nonumber \\
&=& F( (\ket{0_x^L}\bra{0_x^L}+\ket{1_x^L}\bra{1_x^L})_A\otimes(\ket{0_x^L}\bra{0_x^L}+\ket{1_x^L}\bra{1_x^L})_B,  \nonumber \\
&& \;\;\;\;(\ket{0_y^L}\bra{0_y^L}+\ket{1_y^L}\bra{1_y^L})_A \otimes(\ket{0_y^L}\bra{0_y^L}+\ket{1_y^L}\bra{1_y^L})_B  )    \nonumber \\ 
& =   &  F(\ket{0_x^L}\bra{0_x^L}+\ket{1_x^L}\bra{1_x^L}, \ket{0_y^L}\bra{0_y^L}+\ket{1_y^L}\bra{1_y^L}) ^2
\end{eqnarray}

In Ref.~\cite{cao2015discrete}, it was shown that
\begin{eqnarray}
&&F(\ket{0_x^L}\bra{0_x^L}+\ket{1_x^L}\bra{1_x^L}, \ket{0_y^L}\bra{0_y^L}+\ket{1_y^L}\bra{1_y^L})  \\
&\ge &\left| \frac{\sum_{l=0}^{\infty} \frac{\mu^{lN+j}}{(lN+j)!} 2^{-\frac{lN+j}{2}} \left(\cos\frac{lN+j}{4}\pi+\sin\frac{lN+j}{4}\pi\right) }{\sum_{l=0}^{\infty}\frac{\mu^{lN+j}}{(lN+j)!}} \right|.   \nonumber
\end{eqnarray}

Thus Eq.~\eqref{eq:fidelity} in the main text holds.

In addition, in Ref.~\cite{cao2015discrete}, it was shown that 
\begin{eqnarray}
& &\left| \frac{\sum_{l=0}^{\infty} \frac{\mu^{lN+j}}{(lN+j)!} 2^{-\frac{lN+j}{2}} \left(\cos\frac{lN+j}{4}\pi+\sin\frac{lN+j}{4}\pi\right) }{\sum_{l=0}^{\infty}\frac{\mu^{lN+j}}{(lN+j)!}} \right|   \nonumber \\
& \ge & 1 -\left(1-2^{-\frac{N}{2}} \cos\frac{N}{4}\pi \right)\frac{\mu^N}{(N+1)!}.
\end{eqnarray}
So we have
\begin{eqnarray}
&&F_{j,j}(\rho_{AB}^X, \rho_{AB}^Y)  \nonumber \\
&=& \left| \frac{\sum_{l=0}^{\infty} \frac{\mu^{lN+j}}{(lN+j)!} 2^{-\frac{lN+j}{2}} \left(\cos\frac{lN+j}{4}\pi+\sin\frac{lN+j}{4}\pi\right) }{\sum_{l=0}^{\infty}\frac{\mu^{lN+j}}{(lN+j)!}} \right|^2  \nonumber \\
& \ge   &  \left(1 -\left(1-2^{-\frac{N}{2}} \cos\frac{N}{4}\pi \right)\frac{\mu^N}{(N+1)!} \right)^2 \\
& \ge & 1 - 2\left(1-2^{-\frac{N}{2}} \cos\frac{N}{4}\pi \right)\frac{\mu^N}{(N+1)!}. \nonumber
\end{eqnarray}
Hence, the first-order approximation of the fidelity in the main text is proved.

\section{Decoy-State Parameter Deviation}
\label{AppSec:DecoyBound}

In this section, we show the details on the deviation of decoy state gain and error rate.
Like the previous section, here we will also utilizes some results from Ref.~\cite{cao2015discrete}.

By the quantum coin idea~\cite{GLLP:2004}, we have
\begin{eqnarray} 
&  \sqrt{Y_{i,j}^{\alpha,\mu} Y_{i,j}^{\alpha,\nu} } + \sqrt{ (1-Y_{i,j}^{\alpha,\mu})(1- Y_{i,j}^{\alpha,\nu} ) }  \nonumber    \\
& \ge F(\ket{\lambda_i^\alpha}\ket{\lambda_j^\mu}, \ket{\lambda_i^\alpha}\ket{\lambda_j^\nu}) , \nonumber \\
& \sqrt{Y_{i,j}^{\alpha,\mu}e_{i,j}^{\alpha,\mu} Y_{i,j}^{\alpha,\nu}e_{i,j}^{\alpha,\nu} } + \sqrt{ (1-Y_{i,j}^{\alpha,\mu}e_{i,j}^{\alpha,\mu})(1- Y_{i,j}^{\alpha,\nu}e_{i,j}^{\alpha,\nu} ) }  \nonumber    \\
 & \ge F(\ket{\lambda_i^\alpha}\ket{\lambda_j^\mu}, \ket{\lambda_i^\alpha}\ket{\lambda_j^\nu}) .
\end{eqnarray} 

The right-hand side can be simplified as
\begin{equation}
 F(\ket{\lambda_i^\alpha}\ket{\lambda_j^\mu}, \ket{\lambda_i^\alpha}\ket{\lambda_j^\nu}) =  F(\ket{\lambda_j^\mu}, \ket{\lambda_j^\nu})  \ge F_{\mu\nu}.
\end{equation}
The first inequality is because the first systems of the two states are identical, and the second inequality was shown in Ref.~\cite{cao2015discrete}.

Hence 
\begin{eqnarray} 
&\sqrt{Y_{i,j}^{\alpha,\mu} Y_{i,j}^{\alpha,\nu} } + \sqrt{ (1-Y_{i,j}^{\alpha,\mu})(1- Y_{i,j}^{\alpha,\nu} ) }   \ge F_{\mu\nu} , \nonumber \\
& \sqrt{Y_{i,j}^{\alpha,\mu}e_{i,j}^{\alpha,\mu} Y_{i,j}^{\alpha,\nu}e_{i,j}^{\alpha,\nu} } + \sqrt{ (1-Y_{i,j}^{\alpha,\mu}e_{i,j}^{\alpha,\mu})(1- Y_{i,j}^{\alpha,\nu}e_{i,j}^{\alpha,\nu} ) } \nonumber \\
&   \ge  F_{\mu\nu} .
\end{eqnarray} 

In Ref.~\cite{cao2015discrete}, it was shown that
if 
\begin{equation}
\sqrt{xy} + \sqrt{(1-x)(1-y)} \ge  F_{\mu \nu},
\end{equation}
then 
\begin{equation}
| x- y|  \le  \sqrt {1- F_{\mu \nu}^2}.
\end{equation}

Hence we have
\begin{eqnarray}
|Y_{i,j}^{\alpha,\mu}-Y_{i,j}^{\alpha,\nu}|   & \le   \sqrt{1- F^{2}_{\mu\nu}} , \\
|Y_{i,j}^{\alpha,\mu}e_{i,j}^{\alpha,\mu}-Y_{i,j}^{\alpha,\nu}e_{i,j}^{\alpha,\nu}|   & \le  \sqrt{1- F^{2}_{\mu\nu}}   . \nonumber 
\end{eqnarray} 
This finishes the proof.

\section{Simulation}
\label{AppSec:Sim}

In this section, we describe our simulation model and calculate the key rate. 

In the simulation model, we have
\begin{eqnarray} \label{App:SimulationModel}
\eta  &=& 10^{ - \alpha_1 L/10} \eta_1 \nonumber \\
Q_{\mu\nu} &=&(Y_0+1-e^{-\eta\mu})(Y_0+1-e^{-\eta\nu}),   \\
E_{\mu\nu} Q_{\mu\nu} &=& Y_0(Y_0 + 2 - e^{-\eta\mu} - e^{-\eta\nu})/2  \nonumber \\
&&+e_d(1-e^{-\eta\mu})(1-e^{-\eta\nu}), \nonumber
\end{eqnarray}
where $L$ is the transmission distance, $\eta$ is the total transmission loss.
For simplicity, we use three states on each side, namely the signal state, decoy state, and vacuum state, denoted as 1,2,3 on Alice's side, and 4,5,6 on Bob's side.

The simulation parameters are as follows: The fiber loss is $\alpha_1 = 0.2~\textrm{db/km}$. Other losses excluding the fibre loss is $\eta_1 = 0.045$. The misalignment error rate is $e_d = 0.033$. The error correction efficiency is $f=1.16$. The dark count is $Y_0 = 1.7 \times 10^{-6}  $.

To estimate the gain and the error rate, we exploit Eqs.~\eqref{Discrete:QESD} to \eqref{eq:finalestimate} in the main text. Then the intensities of the signal state and the decoy state are optimized to maximize the key rate.

\bibliographystyle{apsrev4-1}

\bibliography{BibliDiscrete}

\begin{thebibliography}{14}%
\makeatletter
\providecommand \@ifxundefined [1]{%
 \@ifx{#1\undefined}
}%
\providecommand \@ifnum [1]{%
 \ifnum #1\expandafter \@firstoftwo
 \else \expandafter \@secondoftwo
 \fi
}%
\providecommand \@ifx [1]{%
 \ifx #1\expandafter \@firstoftwo
 \else \expandafter \@secondoftwo
 \fi
}%
\providecommand \natexlab [1]{#1}%
\providecommand \enquote  [1]{``#1''}%
\providecommand \bibnamefont  [1]{#1}%
\providecommand \bibfnamefont [1]{#1}%
\providecommand \citenamefont [1]{#1}%
\providecommand \href@noop [0]{\@secondoftwo}%
\providecommand \href [0]{\begingroup \@sanitize@url \@href}%
\providecommand \@href[1]{\@@startlink{#1}\@@href}%
\providecommand \@@href[1]{\endgroup#1\@@endlink}%
\providecommand \@sanitize@url [0]{\catcode `\\12\catcode `\$12\catcode
  `\&12\catcode `\#12\catcode `\^12\catcode `\_12\catcode `\%12\relax}%
\providecommand \@@startlink[1]{}%
\providecommand \@@endlink[0]{}%
\providecommand \url  [0]{\begingroup\@sanitize@url \@url }%
\providecommand \@url [1]{\endgroup\@href {#1}{\urlprefix }}%
\providecommand \urlprefix  [0]{URL }%
\providecommand \Eprint [0]{\href }%
\providecommand \doibase [0]{http://dx.doi.org/}%
\providecommand \selectlanguage [0]{\@gobble}%
\providecommand \bibinfo  [0]{\@secondoftwo}%
\providecommand \bibfield  [0]{\@secondoftwo}%
\providecommand \translation [1]{[#1]}%
\providecommand \BibitemOpen [0]{}%
\providecommand \bibitemStop [0]{}%
\providecommand \bibitemNoStop [0]{.\EOS\space}%
\providecommand \EOS [0]{\spacefactor3000\relax}%
\providecommand \BibitemShut  [1]{\csname bibitem#1\endcsname}%
\let\auto@bib@innerbib\@empty
\bibitem [{\citenamefont {Bennett}\ and\ \citenamefont
  {Brassard}(1984)}]{Bennett:BB84:1984}%
  \BibitemOpen
  \bibfield  {author} {\bibinfo {author} {\bibfnamefont {C.~H.}\ \bibnamefont
  {Bennett}}\ and\ \bibinfo {author} {\bibfnamefont {G.}~\bibnamefont
  {Brassard}},\ }in\ \href@noop {} {\emph {\bibinfo {booktitle} {Proceedings of
  the IEEE International Conference on Computers, Systems and Signal
  Processing}}}\ (\bibinfo  {publisher} {IEEE},\ \bibinfo {address} {New
  York},\ \bibinfo {year} {1984})\ pp.\ \bibinfo {pages} {175--179}\BibitemShut
  {NoStop}%
\bibitem [{\citenamefont {Lo}\ \emph {et~al.}(2012)\citenamefont {Lo},
  \citenamefont {Curty},\ and\ \citenamefont {Qi}}]{Lo:MDIQKD:2012}%
  \BibitemOpen
  \bibfield  {author} {\bibinfo {author} {\bibfnamefont {H.-K.}\ \bibnamefont
  {Lo}}, \bibinfo {author} {\bibfnamefont {M.}~\bibnamefont {Curty}}, \ and\
  \bibinfo {author} {\bibfnamefont {B.}~\bibnamefont {Qi}},\ }\href {\doibase
  10.1103/PhysRevLett.108.130503} {\bibfield  {journal} {\bibinfo  {journal}
  {Phys. Rev. Lett.}\ }\textbf {\bibinfo {volume} {108}},\ \bibinfo {pages}
  {130503} (\bibinfo {year} {2012})}\BibitemShut {NoStop}%
\bibitem [{\citenamefont {Ekert}(1991)}]{Ekert:QKD:1991}%
  \BibitemOpen
  \bibfield  {author} {\bibinfo {author} {\bibfnamefont {A.~K.}\ \bibnamefont
  {Ekert}},\ }\href@noop {} {\bibfield  {journal} {\bibinfo  {journal} {Phys.
  Rev. Lett.}\ }\textbf {\bibinfo {volume} {67}},\ \bibinfo {pages} {661}
  (\bibinfo {year} {1991})}\BibitemShut {NoStop}%
\bibitem [{\citenamefont {Xu}\ \emph {et~al.}(2012)\citenamefont {Xu},
  \citenamefont {Qi}, \citenamefont {Ma}, \citenamefont {Xu}, \citenamefont
  {Zheng},\ and\ \citenamefont {Lo}}]{Xu:QRNG:2012}%
  \BibitemOpen
  \bibfield  {author} {\bibinfo {author} {\bibfnamefont {F.}~\bibnamefont
  {Xu}}, \bibinfo {author} {\bibfnamefont {B.}~\bibnamefont {Qi}}, \bibinfo
  {author} {\bibfnamefont {X.}~\bibnamefont {Ma}}, \bibinfo {author}
  {\bibfnamefont {H.}~\bibnamefont {Xu}}, \bibinfo {author} {\bibfnamefont
  {H.}~\bibnamefont {Zheng}}, \ and\ \bibinfo {author} {\bibfnamefont {H.-K.}\
  \bibnamefont {Lo}},\ }\href@noop {} {\bibfield  {journal} {\bibinfo
  {journal} {Opt. Express}\ }\textbf {\bibinfo {volume} {20}},\ \bibinfo
  {pages} {12366} (\bibinfo {year} {2012})}\BibitemShut {NoStop}%
\bibitem [{\citenamefont {Cao}\ \emph {et~al.}(2015)\citenamefont {Cao},
  \citenamefont {Zhang}, \citenamefont {Lo},\ and\ \citenamefont
  {Ma}}]{cao2015discrete}%
  \BibitemOpen
  \bibfield  {author} {\bibinfo {author} {\bibfnamefont {Z.}~\bibnamefont
  {Cao}}, \bibinfo {author} {\bibfnamefont {Z.}~\bibnamefont {Zhang}}, \bibinfo
  {author} {\bibfnamefont {H.-K.}\ \bibnamefont {Lo}}, \ and\ \bibinfo {author}
  {\bibfnamefont {X.}~\bibnamefont {Ma}},\ }\href@noop {} {\bibfield  {journal}
  {\bibinfo  {journal} {New J. Phys.}\ }\textbf {\bibinfo {volume} {17}},\
  \bibinfo {pages} {053014} (\bibinfo {year} {2015})}\BibitemShut {NoStop}%
\bibitem [{\citenamefont {Hwang}(2003)}]{Hwang:Decoy:2003}%
  \BibitemOpen
  \bibfield  {author} {\bibinfo {author} {\bibfnamefont {W.-Y.}\ \bibnamefont
  {Hwang}},\ }\href@noop {} {\bibfield  {journal} {\bibinfo  {journal}
  {Phys.~Rev.~Lett.~}\ }\textbf {\bibinfo {volume} {91}},\ \bibinfo {pages}
  {057901} (\bibinfo {year} {2003})}\BibitemShut {NoStop}%
\bibitem [{\citenamefont {Lo}\ \emph {et~al.}(2005)\citenamefont {Lo},
  \citenamefont {Ma},\ and\ \citenamefont {Chen}}]{Lo:Decoy:2005}%
  \BibitemOpen
  \bibfield  {author} {\bibinfo {author} {\bibfnamefont {H.-K.}\ \bibnamefont
  {Lo}}, \bibinfo {author} {\bibfnamefont {X.}~\bibnamefont {Ma}}, \ and\
  \bibinfo {author} {\bibfnamefont {K.}~\bibnamefont {Chen}},\ }\href@noop {}
  {\bibfield  {journal} {\bibinfo  {journal} {Phys.~Rev.~Lett.~}\ }\textbf
  {\bibinfo {volume} {94}},\ \bibinfo {pages} {230504} (\bibinfo {year}
  {2005})}\BibitemShut {NoStop}%
\bibitem [{\citenamefont {Wang}(2005)}]{Wang:Decoy:2005}%
  \BibitemOpen
  \bibfield  {author} {\bibinfo {author} {\bibfnamefont {X.-B.}\ \bibnamefont
  {Wang}},\ }\href@noop {} {\bibfield  {journal} {\bibinfo  {journal}
  {Phys.~Rev.~Lett.~}\ }\textbf {\bibinfo {volume} {94}},\ \bibinfo {pages}
  {230503} (\bibinfo {year} {2005})}\BibitemShut {NoStop}%
\bibitem [{\citenamefont {Tamaki}\ \emph {et~al.}(2012)\citenamefont {Tamaki},
  \citenamefont {Lo}, \citenamefont {Fung},\ and\ \citenamefont
  {Qi}}]{tamaki2012phase}%
  \BibitemOpen
  \bibfield  {author} {\bibinfo {author} {\bibfnamefont {K.}~\bibnamefont
  {Tamaki}}, \bibinfo {author} {\bibfnamefont {H.-K.}\ \bibnamefont {Lo}},
  \bibinfo {author} {\bibfnamefont {C.-H.~F.}\ \bibnamefont {Fung}}, \ and\
  \bibinfo {author} {\bibfnamefont {B.}~\bibnamefont {Qi}},\ }\href@noop {}
  {\bibfield  {journal} {\bibinfo  {journal} {Phys. Rev. A}\ }\textbf {\bibinfo
  {volume} {85}},\ \bibinfo {pages} {042307} (\bibinfo {year}
  {2012})}\BibitemShut {NoStop}%
\bibitem [{\citenamefont {Tang}\ \emph {et~al.}(2013)\citenamefont {Tang},
  \citenamefont {Yin}, \citenamefont {Ma}, \citenamefont {Fung}, \citenamefont
  {Liu}, \citenamefont {Yong}, \citenamefont {Chen}, \citenamefont {Peng},
  \citenamefont {Chen},\ and\ \citenamefont {Pan}}]{tang2013source}%
  \BibitemOpen
  \bibfield  {author} {\bibinfo {author} {\bibfnamefont {Y.-L.}\ \bibnamefont
  {Tang}}, \bibinfo {author} {\bibfnamefont {H.-L.}\ \bibnamefont {Yin}},
  \bibinfo {author} {\bibfnamefont {X.}~\bibnamefont {Ma}}, \bibinfo {author}
  {\bibfnamefont {C.-H.~F.}\ \bibnamefont {Fung}}, \bibinfo {author}
  {\bibfnamefont {Y.}~\bibnamefont {Liu}}, \bibinfo {author} {\bibfnamefont
  {H.-L.}\ \bibnamefont {Yong}}, \bibinfo {author} {\bibfnamefont {T.-Y.}\
  \bibnamefont {Chen}}, \bibinfo {author} {\bibfnamefont {C.-Z.}\ \bibnamefont
  {Peng}}, \bibinfo {author} {\bibfnamefont {Z.-B.}\ \bibnamefont {Chen}}, \
  and\ \bibinfo {author} {\bibfnamefont {J.-W.}\ \bibnamefont {Pan}},\
  }\href@noop {} {\bibfield  {journal} {\bibinfo  {journal} {Phys. Rev. A}\
  }\textbf {\bibinfo {volume} {88}},\ \bibinfo {pages} {022308} (\bibinfo
  {year} {2013})}\BibitemShut {NoStop}%
\bibitem [{\citenamefont {Glauber}(1963)}]{PhysRev.131.2766}%
  \BibitemOpen
  \bibfield  {author} {\bibinfo {author} {\bibfnamefont {R.~J.}\ \bibnamefont
  {Glauber}},\ }\href@noop {} {\bibfield  {journal} {\bibinfo  {journal} {Phys.
  Rev.}\ }\textbf {\bibinfo {volume} {131}},\ \bibinfo {pages} {2766} (\bibinfo
  {year} {1963})}\BibitemShut {NoStop}%
\bibitem [{\citenamefont {Lo}\ and\ \citenamefont
  {Preskill}(2007)}]{LoPreskill:NonRan:2007}%
  \BibitemOpen
  \bibfield  {author} {\bibinfo {author} {\bibfnamefont {H.-K.}\ \bibnamefont
  {Lo}}\ and\ \bibinfo {author} {\bibfnamefont {J.}~\bibnamefont {Preskill}},\
  }\href@noop {} {\bibfield  {journal} {\bibinfo  {journal} {Quantum
  Inf.~Comput.}\ }\textbf {\bibinfo {volume} {7}},\ \bibinfo {pages} {0431}
  (\bibinfo {year} {2007})}\BibitemShut {NoStop}%
\bibitem [{\citenamefont {Branciard}\ \emph {et~al.}(2013)\citenamefont
  {Branciard}, \citenamefont {Rosset}, \citenamefont {Liang},\ and\
  \citenamefont {Gisin}}]{branciard2013measurement}%
  \BibitemOpen
  \bibfield  {author} {\bibinfo {author} {\bibfnamefont {C.}~\bibnamefont
  {Branciard}}, \bibinfo {author} {\bibfnamefont {D.}~\bibnamefont {Rosset}},
  \bibinfo {author} {\bibfnamefont {Y.-C.}\ \bibnamefont {Liang}}, \ and\
  \bibinfo {author} {\bibfnamefont {N.}~\bibnamefont {Gisin}},\ }\href@noop {}
  {\bibfield  {journal} {\bibinfo  {journal} {Phys. Rev. Lett.}\ }\textbf
  {\bibinfo {volume} {110}},\ \bibinfo {pages} {060405} (\bibinfo {year}
  {2013})}\BibitemShut {NoStop}%
\bibitem [{\citenamefont {Gottesman}\ \emph {et~al.}(2004)\citenamefont
  {Gottesman}, \citenamefont {Lo}, \citenamefont {L\"utkenhaus},\ and\
  \citenamefont {Preskill}}]{GLLP:2004}%
  \BibitemOpen
  \bibfield  {author} {\bibinfo {author} {\bibfnamefont {D.}~\bibnamefont
  {Gottesman}}, \bibinfo {author} {\bibfnamefont {H.-K.}\ \bibnamefont {Lo}},
  \bibinfo {author} {\bibfnamefont {N.}~\bibnamefont {L\"utkenhaus}}, \ and\
  \bibinfo {author} {\bibfnamefont {J.}~\bibnamefont {Preskill}},\ }\href@noop
  {} {\bibfield  {journal} {\bibinfo  {journal} {Quantum Inf.~Comput.}\
  }\textbf {\bibinfo {volume} {4}},\ \bibinfo {pages} {325} (\bibinfo {year}
  {2004})}\BibitemShut {NoStop}%
\end{thebibliography}%

\end{document}